\documentclass{llncs}
\usepackage{graphics}
\usepackage{mathptmx}       
\usepackage{helvet}         
\usepackage{courier}        
\usepackage{type1cm}        
\usepackage{makeidx}         
\usepackage{graphicx}        
\usepackage{multicol}        
\usepackage[bottom]{footmisc}
\usepackage{subfigure}
\usepackage{amsfonts}
\usepackage[cmex10]{amsmath}

\newcommand{\comment}[1]{}
\begin{document}

\title{On the breakup patterns of urban networks under load}

\author{Marco Cogoni \and Giovanni Busonera}
\institute{CRS4 - 09010 Pula (CA), Italy\\
\email{marco.cogoni@crs4.it}
}

\maketitle

\section{Introduction}
Traffic has been extensively studied in recent years with a focus on the free-flow to congestion transition. Several models have been employed, from microscopic to coarse-grained approaches~\cite{Helbing2001}. Very recently, a new perspective, based on percolation~\cite{Havlin2015}, has been proposed to study traffic flows in large cities with real GPS data. This approach disregards vehicle dynamics, by focusing on the ability of each road to guarantee transportation efficiency above some minimum threshold. What emerged from these studies was that, beyond the well known passage from free-flow to the congested state, a percolation transition exists: when observing the network as a whole, it progressively decomposes from a single giant (strongly connected) component to a set of separated clusters, each able to sustain traffic within its boundaries above some threshold speed, but functionally disconnected from the others~\cite{Havlin2015}.
It is known that a urban network graph undergoes a critical percolation transition when a fraction of its edges is removed. The resulting strongly connected components form a structure of clusters whose size distribution follows a power law with critical exponent $\tau$~\cite{Havlin2015}. This exponent changes under different traffic regimes~\cite{Havlin2019}. A complete explanation of which factors affect $\tau$ is still lacking. We argue that spatial correlations may play a major role influencing the $\tau$ behavior since their range is known to grow when going from free-flow to congested real traffic~\cite{Rempe2016}. We show that values of $\tau$ for random noise with increasing spatial correlations and those from GPS data for increasing congestion follow a similar behavior.

\section{Methods}
We obtained the transportation networks (only roads open to cars) from OpenStreetMap in the form of directed weighted graphs centered on New York City and London ($1600$ km$^2$ each). 
To perform random percolation, edges are uniformly removed whereas, for real data, they are deleted when the local average speed is under a critical threshold (congested). In both cases, we localize the critical probability threshold $p_c$, which determines the average number of remaining edges, by performing a set of percolation instances for $p\in(0,1)$ and choosing $p$ that leads to the maximum size of the second largest cluster.
In the synthetic, uncorrelated case we use a uniform random generator in $(0,1)$ to assign a "speed" value to each edge, while to induce spatial correlation, we compute the graph Laplacian and its spectrum via a fast approximated method \cite{PyGSP2017}, then we perform a Graph Fourier Transform of the uncorrelated noise and filter it~\cite{Prakash1992} by multiplying the eigenvalues with a power-law $f(q) \sim q^{-\lambda}$: a rapid decay ($\lambda\rightarrow 2$) leads to long spatial correlations, while $\lambda\rightarrow 0$ is equivalent to uncorrelated noise.
For the real data, we first compute the average speed value on each edge for the relevant time span (UBER data has a one hour granularity) and then divide it by the maximum speed observed on that edge over six months. 
We compared the existing results from Ref.~\cite{Havlin2015} (based on undisclosed data) with our percolation analysis conducted both on real traffic (from the UBER Movement datasets\comment{\footnote{https://movement.uber.com}}) and for synthetic data with increasing spatial correlation.
To compute the critical exponent $\tau$ we perform a linear fit on a log plot of the binned distribution of the cluster sizes for each percolation instance~\cite{Campi2005}.
Finally, to produce a spatial map of the city areas associated to the main clusters, we plot the number of times that, over several percolation replicas, each edge belongs to the 1st, 2nd and 3rd largest cluster, respectively.

\section{Results}
\begin{figure}[htb]
\centering
    \includegraphics[width=0.49\textwidth]{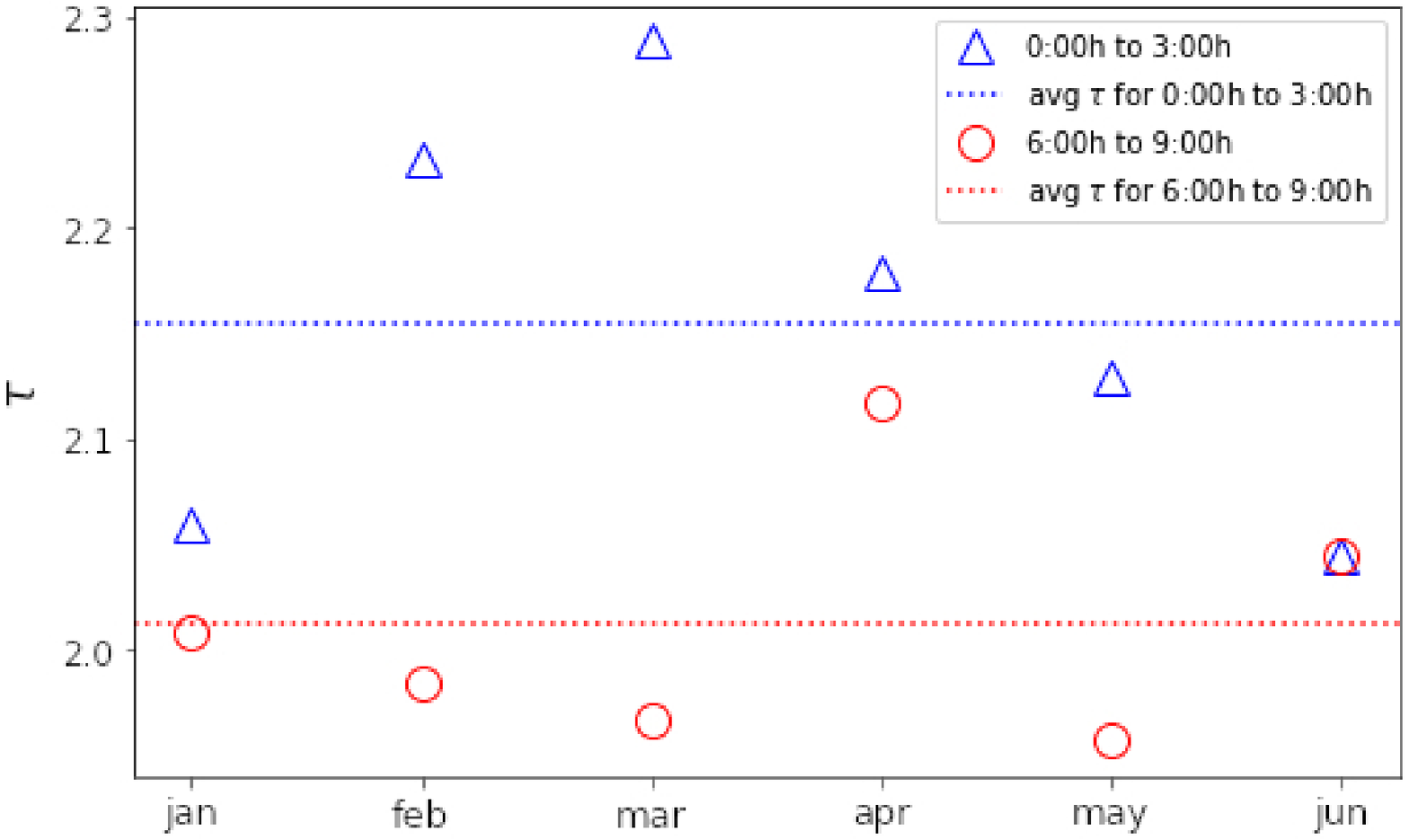}
    \includegraphics[width=0.49\textwidth]{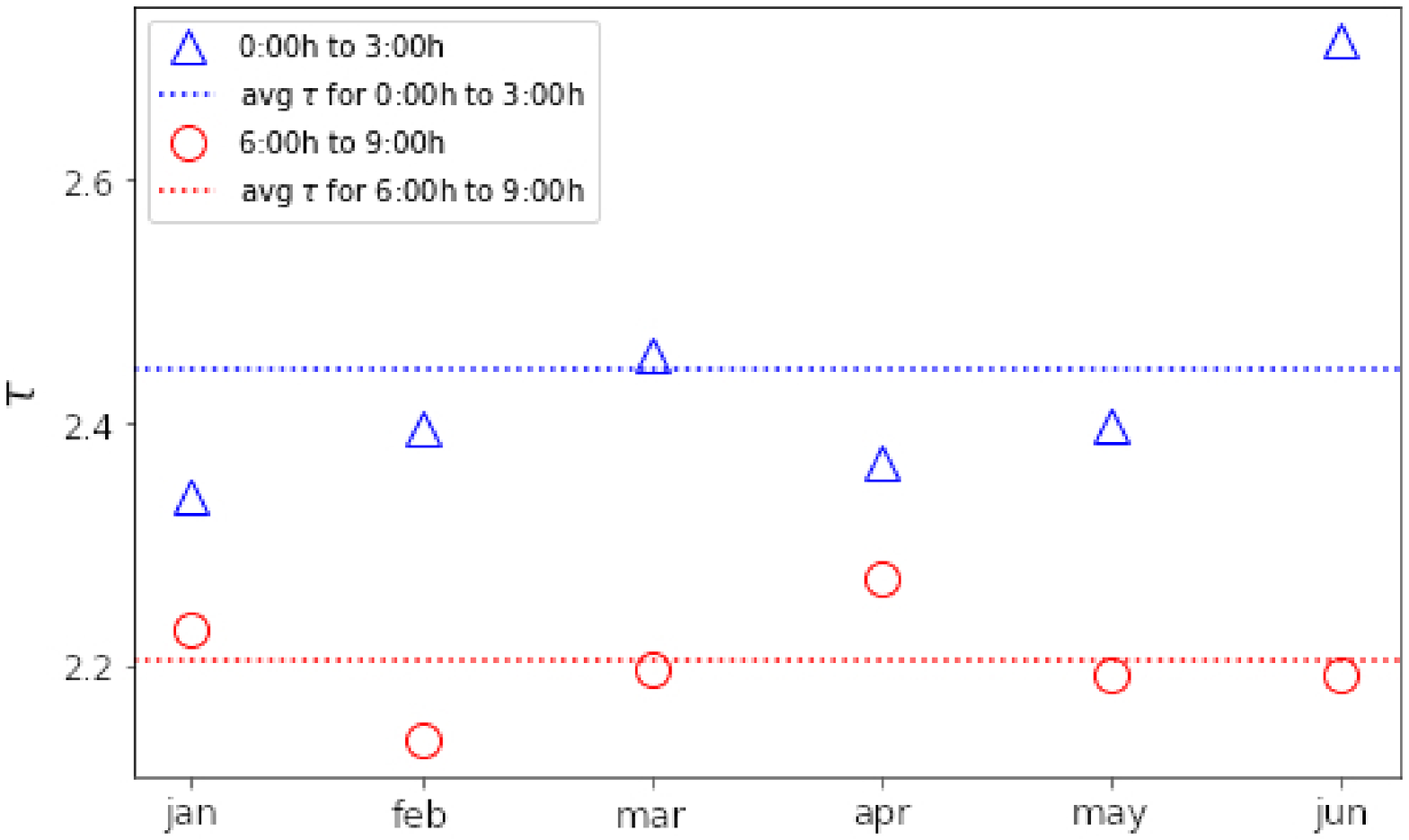}
\caption{Monthly average $\tau$ for London(left) and NYC(right) (rush hours(red) and off-peak(blue))}
\label{fig:tau_exponent}
\end{figure}

We first use real data to perform percolation runs on a monthly basis for London and New York City networks: during low congestion periods (0AM-3AM), $\tau$ is close to the one associated to percolation in mean-field networks ($\tau\sim2.5$), while during rush hours (6AM-9AM) the exponent approaches the one observed for a regular square lattice ($\tau\sim2.05$), as shown in Fig.~\ref{fig:tau_exponent}.
We subsequently analyze random percolation with increasing spatial correlations. For each city, $\tau$ slowly decreases with longer correlations. For both cities $\tau$ ranges from $\sim2.2$ for $\lambda=0$ to $\sim2.0$ for $\lambda\sim2.0$. These results show that a simple random percolation is able to capture the basic properties of real uncongested traffic (higher $\tau$ similar to mean field networks), and that increasing spatial correlations leads to a different cluster size distribution (lower $\tau$ similar to lattice percolation), typical of rush hours. We believe this to be a useful result for a better understanding of how traffic clusters behave over large urban areas under different congestion levels.

\begin{figure}[htb]
\includegraphics[width=0.33\textwidth]{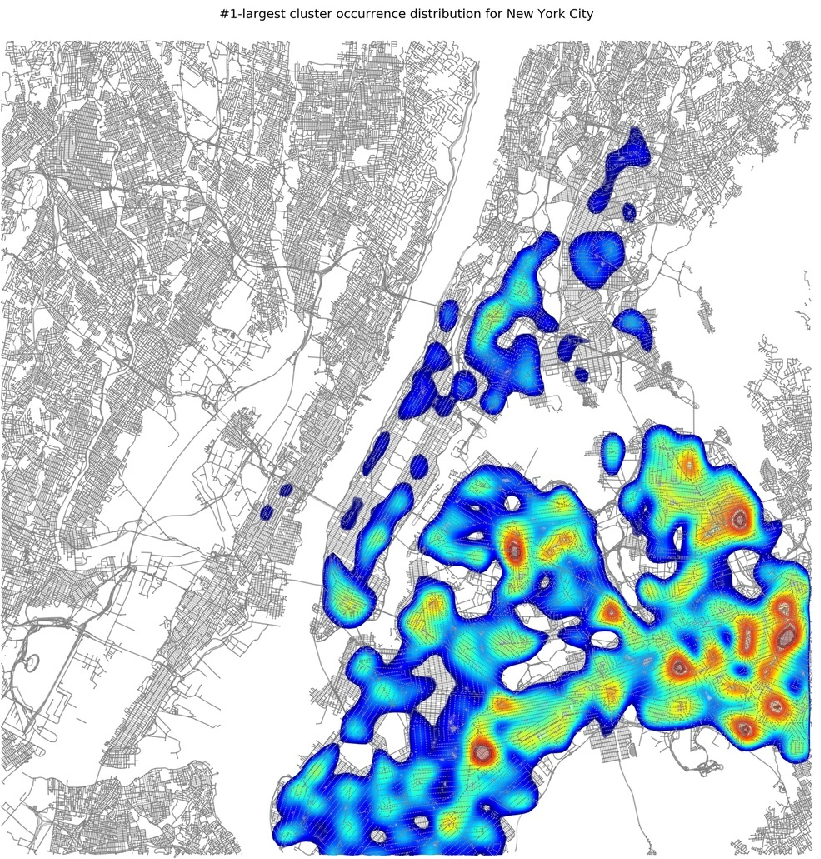}
\includegraphics[width=0.33\textwidth]{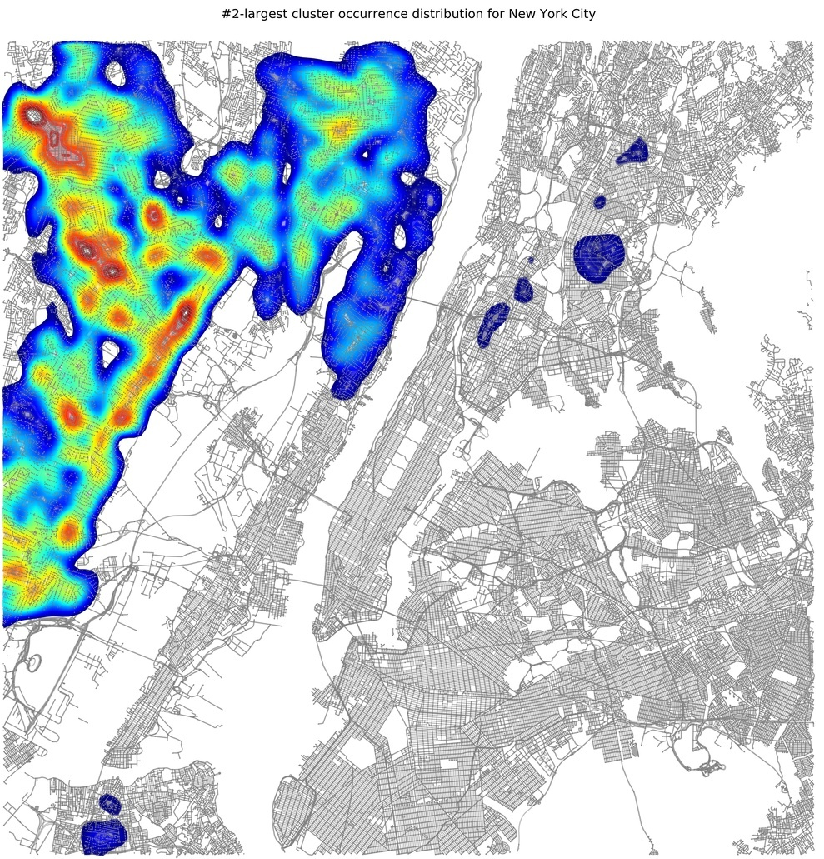}
\includegraphics[width=0.33\textwidth]{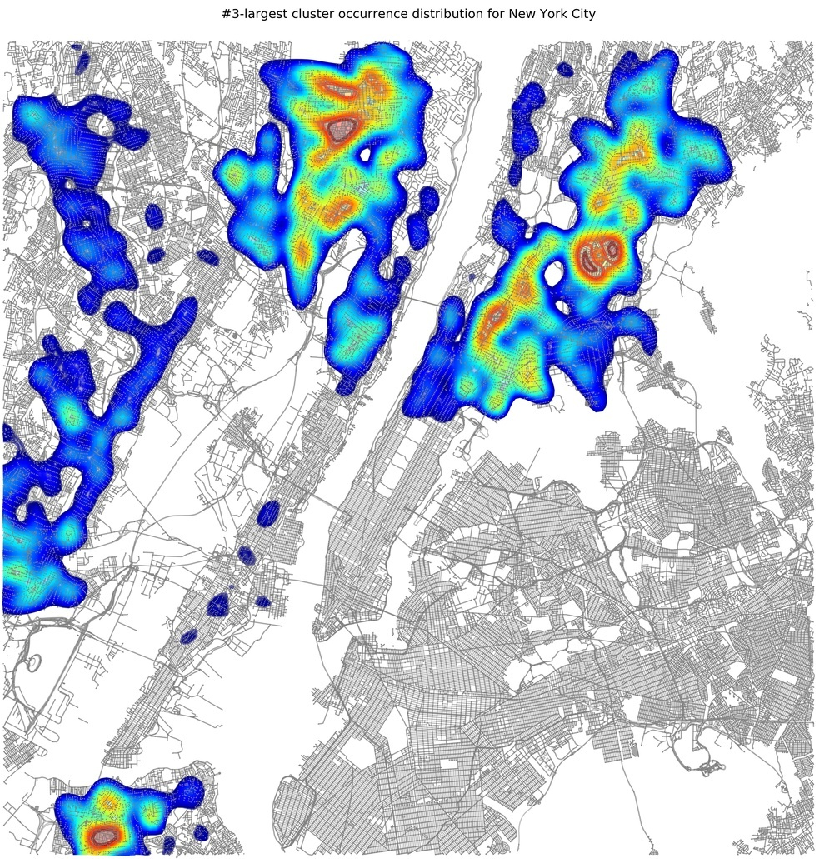}

\includegraphics[width=0.33\textwidth]{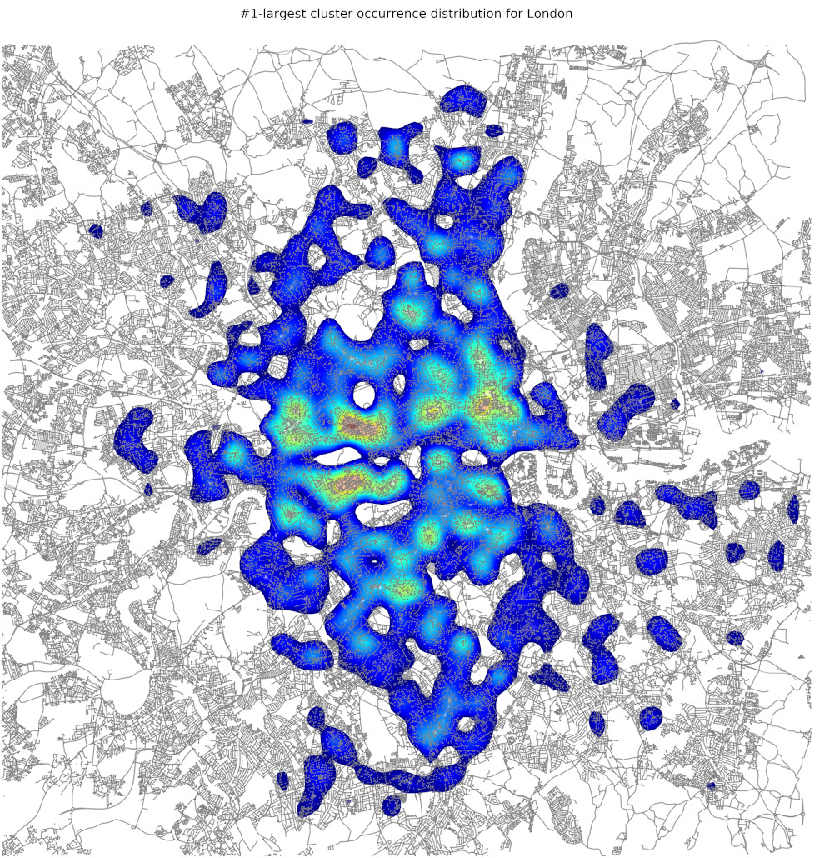}
\includegraphics[width=0.33\textwidth]{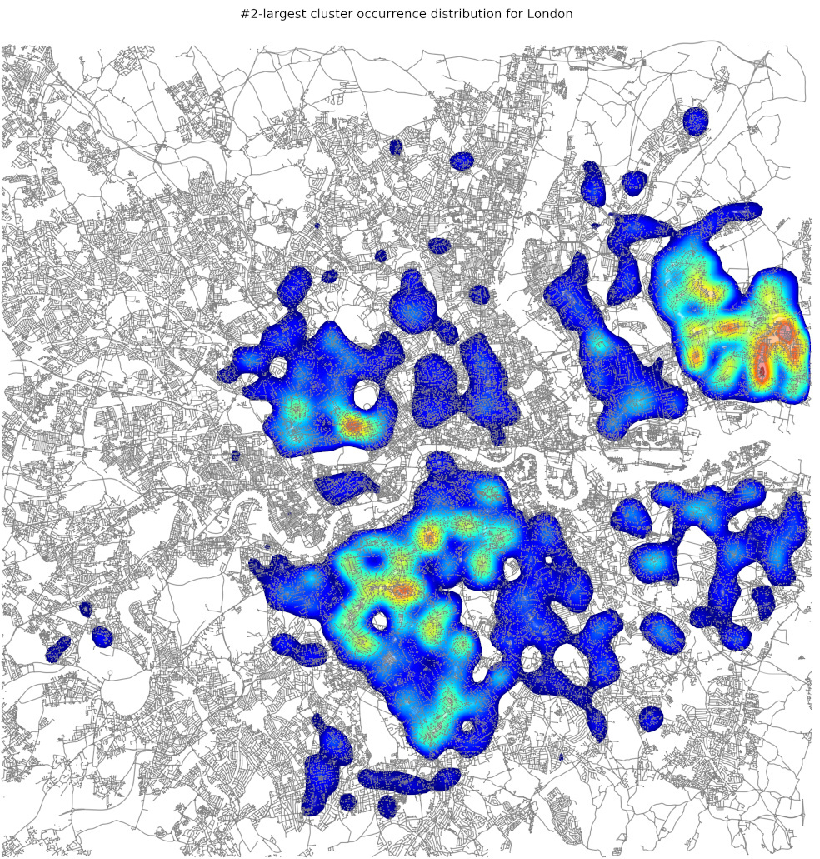}
\includegraphics[width=0.33\textwidth]{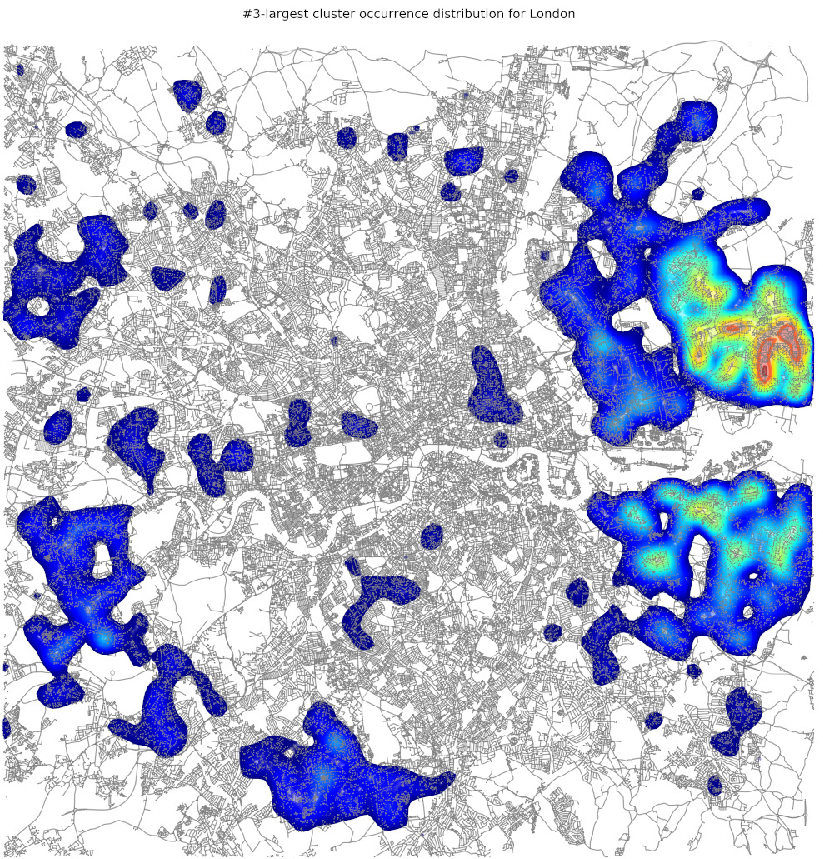}
\caption{From left to right: 1st, 2nd and 3rd largest cluster spatial distributions for NYC (top) and London (bottom). Blue (red) means that an edge was associated to the cluster in $80\%$ ($100\%$) of the replicas.}
\label{fig:3GC}
\end{figure}

We finally present some preliminary evidence that, at criticality, the largest clusters display spatial predictability and that cluster configurations from $2000$ replicas lead to a small number of breakup patterns, specific for each city. This holds both for random percolation and real traffic. Results obtained from synthetic percolation (uncorrelated) are shown in Fig.\ref{fig:3GC} for London and NYC, where the three largest clusters are well localized and spatially distinct. Results from correlated percolation and from real traffic data are qualitatively similar, but the statistics for the latter is limited to one sample per week.
This consistent cluster organization appears to be strongly influenced by local topographical structures such as rivers and bridges.
The maps help visualizing how frequently city areas belong to the main functional traffic clusters and could be useful for city planners to improve city connectivity by easily comparing different road topologies. Moreover, citizens could better choose where to buy a house by selecting an area belonging (on average) to the same efficiently connected cluster as their workplace.

\end{document}